# Modeling of non-planar slicer for improved surface quality in material extrusion 3D printing

**Shadman Tajwar Shahid[1, *]**

[1]Department of Mechanical Engineering, Military Institute of Science and Technology, Mirpur Cantonment, Dhaka-1216, Bangladesh

*Corresponding author email: shadmantajwar@me.mist.ac.bd

**Abstract:** This paper introduces an algorithm to generate a 3D extruder path, combining classic planar and non-planar layers to enhance the surface quality and accuracy of complex 3D printed parts. Material extrusion 3D printing, due to its layer-by-layer construction method, produces parts with a discretization effect—commonly referred to as the "staircase" effect— particularly on near-flat surfaces. The algorithm addresses this issue using a mixed-layer approach that uses 3D non-planar layers for the surfaces that would benefit from non-planar printing, and planar layers for the remaining regions. Existing studies have demonstrated similar combined layer methods, but are often limited in the variety and complexity of shapes they can process due to their inherent slicing techniques. This algorithm presents a universal approach to non-planar extruder path generation by identifying the non-planar surfaces and generating non-planar extruder paths that conform to the object's surface. Subsequently, it identifies the space occupied by the non-planar layers and removes it from the original mesh to produce a collision-free planar-only mesh, sliced using classic planar methods. The algorithm was evaluated on objects of various complex shapes, comparing the results with outputs from standard planar slicers. The improvement in surface accuracy was also quantified by measuring the Chamfer Distance. Specifically, it is shown that the algorithm can generate non-planar extruder paths of complex geometries, improving surface quality.

**Keywords:** Non-planar slicing algorithm, Curved-layer FDM, Complex geometries, Surface accuracy, Extruder path generation

# 1. Introduction

The most common form of additive manufacturing is Fused Filament Fabrication (FFF) or Fused Deposition Modelling (FDM), where heated plastics are deposited on a flat surface and extruded to follow the object's shape. Height is achieved by raising the nozzle after completing each layer and stacking one layer on top of another. Due to the layer-by-layer nature of FDM, it is inherently subject to a discretization effect, called stair-stepping. The stair-stepping effect becomes increasingly apparent as the surfaces approach shallow angles, deteriorating the object's aesthetic and mechanical properties.

Researchers have explored several techniques to improve the surface quality of 3D prints. One common technique is adaptive layer height, which adjusts the height of each layer when approaching shallow surfaces to enhance smoothness. Adaptive layering is the only fully planar technique that improves surface quality. Most adaptive layering techniques use the cusp height limit method introduced by Dolenc et al. [1] to determine the layer height. Kulkarni et al. [2] developed adaptive slicing algorithms for parametric geometry directly from CAD models whereas Sabourin et al. [3] proposed a faceted geometry-based adaptive slicing procedure. However, this method is ineffective when the model has high complexity in the layer stacking direction.

A new research trend focuses on continuously using the three axes to create parts that truly conform to the object's shape, known as non-planar printing. This method leverages the full 3D capabilities of printers to produce surfaces with higher accuracy and reduced support structure and stair-step effects.

Etienne proposed a semi-nonplanar slicing algorithm called CurviSlicer to obtain a better surface finish with Cartesian FDM 3D printers [4]. The algorithm curves the layers that make the print head follow the natural curve of the surface. The algorithm does not directly compute the curved path but optimizes for a deformation of the model, which is sliced using the standard planar approach, and the curved path is obtained by inverse deformation of the planar toolpath. In this case, the layer height varies from 0.1 to 0.6 mm. The printed models produced showed good accuracy and surface quality compared with uniform slicing methods

One of the first implementations of nonplanar printing by FDM was by Chakrabarty et al. [5] called Curved Layer FDM (CLFDM). The method uses a parametric surface to generate the toolpaths based on CNC concepts and create thin-shelled parts. Although the algorithms were presented, real physical objects were not printed. Parts printed with curved layers using faceted geometry were introduced by Singammeni et al. [6] where the curved parts were printed on support structures printed using planar methods. CAM modules of modeling software, suited for CNC machining were used to generate the extruder paths. Standard Triangle Language (STL) format file processing was attempted, but the final test pieces were made using the CAM route. Jin et al. [7] presented a slicing procedure where a tessellated surface is fitted into a B-spline surface. The original surface is offset to obtain the inner layers. The first extruder path is along one of the edges of the surface, and offsetting generates adjacent paths. To avoid the inward offset of the boundary, an enlargement coefficient was applied to the offsetting so that the boundary intersects the object. The slicing approach is specific to their application and not a general approach. Rodriguez et al. [8] proposed a conformal printing method to create structure on surfaces based on the projection of printing trajectory on triangle tessellated nonplanar surfaces. Actual prints of hexagonal, Hilbert, and re-entrant patterns produced on curved surfaces were presented.

Ahlers et al. [9] implemented an algorithm combining planar and non-planar slicing, using curved layers for exterior shallow surfaces and flat layers for the interior of the part to achieve smooth surfaces in a relatively short time. The non-planar surface was shifted along the z-axis and not along the facet normal. So, to compensate for the non-constant layer thickness, the extrusion flow was multiplied by a correction factor. Due to using Cartesian FDM 3D printers, the method is highly dependent on the geometry of the tool head and part geometry. The algorithm was implemented in freely available Slic3r software, making it a practical and accessible approach to non-planar 3D printing. Maria et al. [10] combined classic planar and non-planar layers to print structures onto complex substrates using a 5-axis robotic arm. In this case, both top and bottom non-planar surfaces were printed. Feng et al. [11] implemented a 5-axis system based on a delta printer with a rotating platform for non-supported curved layer 3d printing. A series of conformal surfaces were generated based on surface offset, and an infill toolpath was generated using the geodesic distance as the shortest zigzag path along the edge of each surface. The algorithm was used to print thin-walled half-spherical shells and skull implants.

The review of related literature reveals that most works on non-planar 3D printing focus on thin-shelled, simple geometries with uniform thickness. Moreover, some of the slicing approaches use CAM packages and parametric geometries to generate the extruder paths, and as a result, are case-specific and not a general use scenario. The capability to print arbitrary shapes from universally compatible 3D model formats such as STL is an important feature of a 3D slicing program as it allows for more compatibility and accessibility. Ahlers et al. [9] and Maria et al. [10] implemented non-planar slicing methods using STL models. However, the non-planar surfaces produced by these methods are continuous freeform surfaces that do not have any holes or splits. In the demonstrated 3D prints, either the boundaries of the non-planar surfaces merge with a perpendicular surface, or all points in the non-planar boundary start at the same z-height. The performance is unknown for complex geometries where the non-planar boundary is distributed at varying z-height with a smooth transition to the planar surface, consists of multiple isolated surfaces at different heights, and there are holes or splits in the surface.

Motivated by the limitations found in this literature, this paper focuses on versatile combined planar and non-planar slicing algorithm that can be applied to a wide variety of geometrically complex objects, and as a result a more universal approach. The methods in this study were intended to be used in a Cartesian 3D printer with a rotating print head. However, a 3-axis machine would suffice if the surface is slightly curved. The extruder paths were generated by a Python program and the graphical simulation using Blender.

## 2. Methods

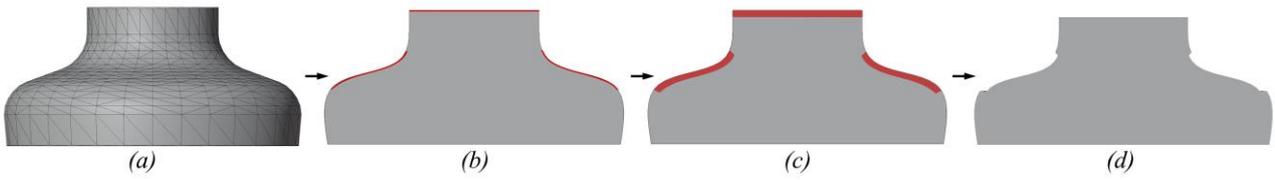

Fig. 1. Overview of non-planar slicing algorithm: a) STL mesh of object to be sliced b) Indentifying non-planar faces c) Generation of non-planar space d) Planar-only mesh after Boolean difference operation.

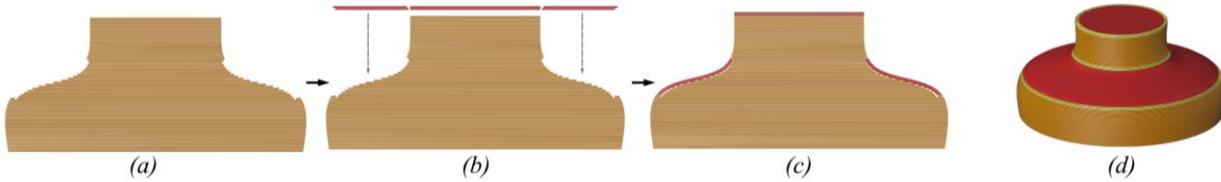

Fig. 2. a) Planar slicing of planar-only mesh b) 2D non-planar layers c) Projection of non-planar layers onto mesh d) 3D view of extruder paths. In orange are the planar walls, in yellow are the non-planar walls, and in red is the non-planar infill.

### 2.1. Overview

This study presents a non-planar slicing method implemented using Python. The visualization of the extruder path was created using Blender. The algorithm is a mixed-layer approach, using curved layers for non-planar surfaces and planar layers for the remaining part of the object.

The object will consist of planar layers, while non-planar layers will make the surfaces that show a high staircase effect. The planar and non-planar layers are formed of one or more outer perimeters of extrusion, called walls, and an inner region, called infill. The planar layers allow the object to be printed in a relatively short time and support the non-planar layers. On the other hand, the non-planar layers improve accuracy and enhance surface quality by eliminating the staircase effect.

### 2.2. Planar Extruder Path Generation

In non-planar 3D printing, the idea is to use non-planar layers only for surfaces that would benefit from non-planar printing, and printing the other regions using classic planar methods. The STL model is sliced into evenly spaced parallel cross-sections to generate the planar layers. Slicing the original model would give a cross-section that overlaps with the non-planar layers. So, to produce a collision-free mesh, the space that is occupied by the non-planar layer is removed from the original model, which is then sliced by the classic planar methods.

### 2.3. Non-planar Triangle Identification

The models for 3D printing are made in Computer-Aided Design (CAD) software, which defines model objects by analytic geometry representation. To prepare a CAD model for 3D printing, it is converted into Standard Triangle Language (STL) format, which defines objects in a discrete geometry representation. This conversion process, known as tessellation, involves discretizing the CAD model's surfaces and solid bodies into a mesh of interconnected triangles [12]. The level of tessellation determines the resolution of the STL file—more triangles result in finer detail and a more accurate representation of the object's shape. The STL format consists of a collection of triangles defined by the 3D coordinates of the vertices of triangles and the corresponding normal of the triangle.

Due to the discrete layer-by-layer nature of material extrusion 3D printing, surfaces almost parallel to the build plate or curved surfaces appear to be stepped or jagged. The smoothness of the printed surface depends on the layer height and surface inclination. Thicker layers result in a more visible staircase effect because the transition between layers is more pronounced. Not every surface will benefit from non-planar printing because near horizontal surfaces are printed with high quality with planar layers. The smoothness of nonplanar layers is higher for the same surface as when printed with planar layers if the extrusion width is smaller than the step size [9], where the step size is the length of the exposed outer wall and infill of the layer. Below a certain inclination, it is beneficial to have the surface printed with non-planar layers rather than planar ones. For a given extrusion width and layer height, the threshold angle of surface inclination is as follows:

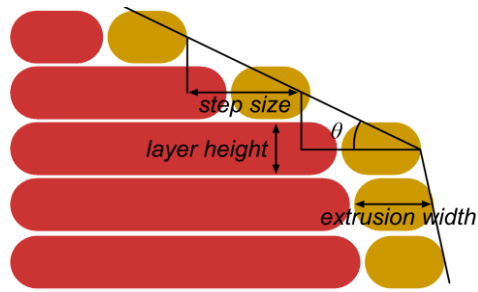

Fig. 3. Effect of surface angle on quality of printed surface.

$$\theta_{th} = tan^{-1} \frac{layer\ height}{extrusion\ width} \quad (1)$$

Testing the triangles for non-planarity is the first step of the algorithm. The triangle's orientation is the triangle facet normal from the STL mesh. If the angle between the triangle facet normal and the print-bed normal is smaller than the threshold angle $\theta_{th}$ and the $k$ component is positive, meaning the triangle is facing upward, then the triangle is non-planar. The surface formed by all such triangles is referred to as the original non-planar surface.

A second test ensures no parts of the object are above the non-planar surface by ray casting from vertices of the triangles vertically upwards. When the ray intersects another triangle, indicating that a part of the object is above the vertex, the triangle is discarded from the non-planar surface. Moreover, a collision test is done by bounding box collision detection, where the bounding box of the collider is according to the shape of the extruder and the target is the original mesh. Proper collision detection is by testing for all points in the extruder paths, this is very accurate but computationally intensive. So, for the tests, collision detection is done only at the vertex point of the triangle, which significantly reduces the number of computations but there is the possibility of collision when the object is complex. The collision test is adequate for a 5-axis 3D printer when the surfaces are distributed at the top of the model and the curvature is not large. The non-planar triangle identification procedure is discussed in the first step of Algorithm 1.

## 2.4. Non-planar Surface Offset

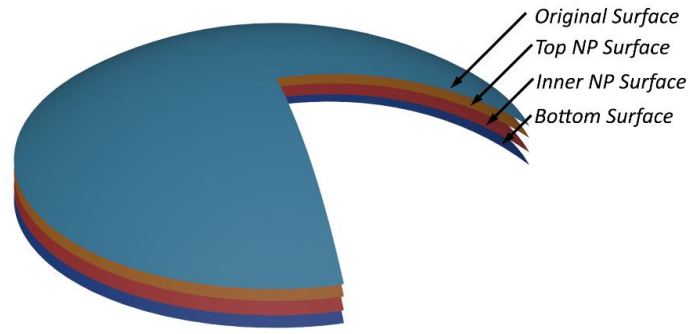

Fig. 4. Different non-planar (NP) surfaces.

Offsetting the original non-planar surface generates the bottom surface of the non-planar space and the surfaces for the top and internal non-planar layers. A constant thickness offsetting method is needed for accuracy and to prevent over and under-filling of non-planar layers. To offset a surface in the STL format, the most direct method is to offset each triangle with the offset distance in its corresponding normal direction. However, this will result in intersections or gaps between the offset surfaces of neighboring triangles. The problem is avoided if the vertices, instead of the triangles are offset. Since the STL file does not contain the vertex normal, normal at each vertex is calculated by using the averaged normal vector method developed by Qu et al. [13]. The vertex normal vector is calculated by averaging the normal of all the triangles connected to the vertex.

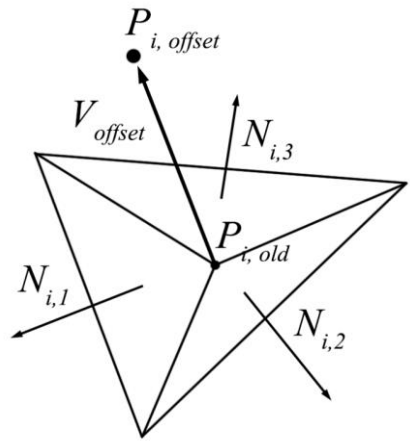

Fig. 5. The normal vertex is calculated from the average of the normal of connected triangles.

$$V_{offset} = \frac{\sum_{j=1}^{n} N_{i,j}}{|\sum_{j=1}^{n} N_{i,j}|} \tag{2}$$

$$P_{i,offset} = P_{i,old} + V_{offset} \times d_{offset} \tag{3}$$

The method works well for vertex located on a relatively smooth surface. However, a vertex located at a sharp corner, such as the corner of a cube, may give errors. Dense and complex meshes with a large surface offset distance can cause self-intersections, looping surfaces, and irregularities. Koc et al. [14] presented a robust method of surface offsetting by using biarcs fitting that addresses these problems. Due to its complexity, it was not currently implemented in the algorithm.

**2.5. Planar-only mesh generation**

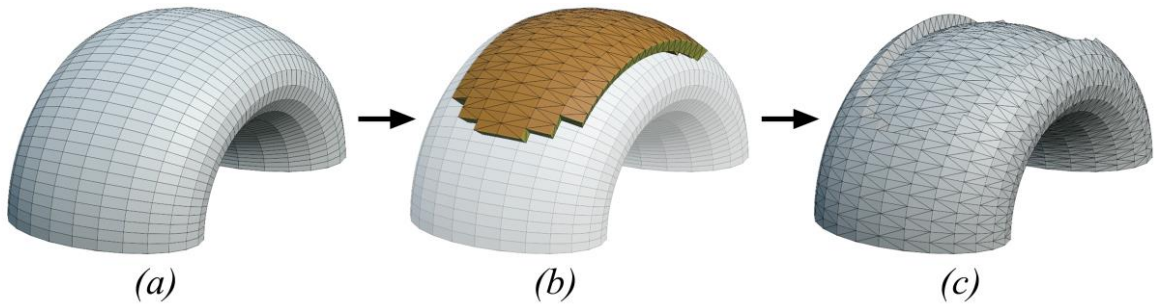

Figure 6. a) Original mesh b) Generation of non-planar space c) Planar-only mesh generation by Boolean difference operation.

To produce planar layers that do not intersect with the non-planar layers, the STL mesh is modified by removing the space occupied by the non-planar layers. The space occupied by the non-planar layers is between the original non-planar surface and the bottom non-planar surface. The offset distance of the bottom surface depends on the layer height and number of non-planar layers. First, the boundaries of the two surfaces are identified, and the gap is filled with new faces by connecting the boundary vertex of the original surface to the same vertex offset to the bottom surface. The original surface, bottom offset surface, and boundary faces are merged to form a unified, closed mesh.

Finally, the newly generated mesh is subtracted from the original mesh using the Boolean difference operation. The Python program used functions from the times and STL libraries. The newly generated mesh contains empty space where the non-planar layers are printed and is referred to as the planar-only mesh.

In the study by Ahlers et al. and Marie et al., the generation of collision-free extruder paths between the non-planar and planar layers is a post-slicing process performed by Boolean subtraction between the 2D cross-section belonging to the planar layer and the layer above. As a result, the outer 3D printed surface cannot simultaneously have both planar and non-planar layers at the same z-height, but only when the planar surface is perpendicular, limiting the complexity of geometry to which the method can be applied. Separating the planar and non-planar parts as separate mesh before the slicing procedure allows for more complex transitions between the planar and non-planar layers.

**2.6. Slicer**

The slicer is a program for 3D printing that transforms STL models into commands for 3D printing machines. At its core, the slicer operates by converting a three-dimensional model into a series of two-dimensional cross-sectional layers [15]. The slicer computes the intersections of the triangles from the STL model with a series of horizontal planes distributed evenly along the z-axis. Each intersection generates a cross-sectional profile that, when layered sequentially, reconstructs the entire 3D object [16]. The object reconstructed through the slicing procedure is an approximation of the STL mesh, whose accuracy depends on the distance between each layer. The smaller the height between each subsequent cutting plane, the more accurate the reconstruction. The input STL model is the planar-only mesh generated after the Boolean difference operation, not the original mesh.

### 2.6.1. Triangle-Plane Intersection

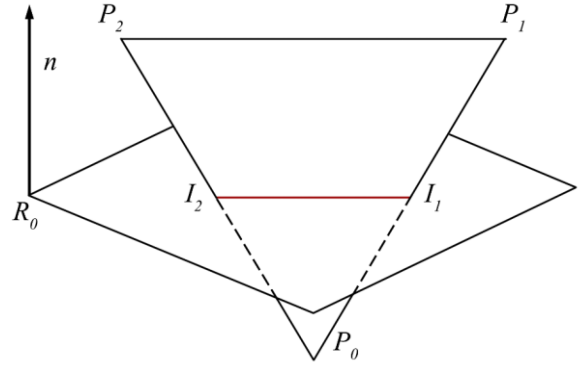

Fig. 7. Triangle-plane intersection.

The 3D coordinates of vertices, faces, and normal are obtained from the STL mesh. Consider a triangle $T$ with vertices $P_0$, $P_1$, and $P_2$ the plane $P$ through point $R_0$ that is parallel to the print bed, with the normal vector pointing along the z-axis, $n = (0, 0, 1)$. When the triangle $T$ does not intersect the plane, all three of the vertices must lie on the same side. On the other hand, when $T$ intersects the plane, one vertices of $T$ must be on the side opposite the other two. The sidedness of the vertices of the edge formed by $P_0$ and $P_1$ is computed by the dot product, $S_0 = n \cdot (R_0 - P_0)$ and $S_1 = n \cdot (R_0 - P_1)$. If $S_0$ and $S_1$ have opposite signs, the vertices $P_0$ and $P_1$ lie on opposite sides of the plane, indicating an intersection [17]. The intersection point of the edge is computed as follows:

$$I_1 = P_0 + t.d \qquad (4)$$

$$t = \frac{n.(R_0 - P_0)}{d.n} \qquad (5)$$

$$d = P_1 - P_0 \qquad (6)$$

Two edges of the triangle $P_0P_1$ and $P_1P_2$ intersect the plane $P$ at $I_1$ and $I_2$ respectively, the intersection of the triangle is the line formed by the two points. When all the z-coordinates of the triangles are the same as the cutting plane height, they are coplanar to the plane, and the intersection is the outer perimeter of the coplanar triangles. The process is repeated for all the triangles, finding out the cross-section of the STL model on the plane at a specific height.

The cross-section is a polygonal chain or multiple chains consisting of straight lines. The coordinates of the straight lines are added in a sequential format, $x\_coords = [x_1, x_2, x_2, x_3, x_3, x_4,...]$ and $y\_coords = [y_1, y_2, y_2, y_3, y_3, y_4,...]$, where each pair of coordinates contains the start and end points of a line segment. The representation is chosen because it simplifies organizing the geometric data. By breaking down the shape into pairs of coordinates representing individual line segments, the format allows for identifying any discontinuities between segments and detecting separate polygonal chains within the cross-section. The polygonal chains produced from slicing cannot be directly used and need to be processed.

**2.6.2. Sorting**

The line segments derived from the cross-section are initially unorganized because the triangles in the original STL file are not arranged in any specific order. Organizing the randomly distributed segments into a clockwise or counter-clockwise sequence is essential for further operations such as generating offsets, infills, and detecting holes.

The sorting algorithm is simple, the direction of the chain is selected, anti-clock or clockwise. The chain direction must be consistent for all the chains in a layer. An arbitrary line segment is selected and the direction is checked and adjusted, for anticlockwise, the line goes from left to right and top to bottom. The end of the line segment is compared with the start and end of each subsequent straight line until a match is found, and a matching line segment is moved to the adjacent, reversing the direction if necessary. The process is repeated until no matches are found, which means either the chain is complete or a discontinuity has been found.

**2.6.3. Hole Identification**

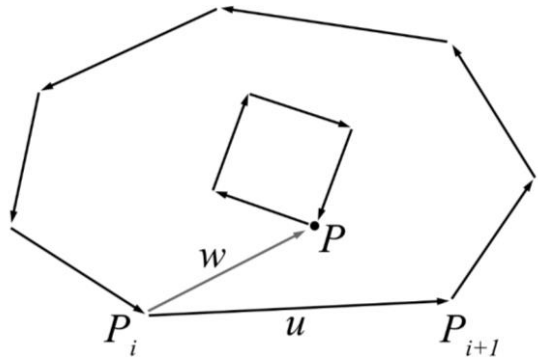

Fig. 8. Hole identification by cross-product.

$$u = P_{i+1} - P_i, w = P - P_i \qquad (7)$$

Holes within the chains are identified by testing the orientation of all vertices in a chain against all line segments of other chains. A sorted polygonal chain is essential at this stage. The orientation of the points depends on the cross-product, $u \times w$. When all points of one chain lie inside another chain, the inner chain is designated a hole corresponding to the outer chain. A hole is considered an inner hole if it lies inside an odd number of outer chains. The inner hole's direction is reversed to that of the outer chain because the walls of the holes are offset in the opposite direction.

## 2.7. Wall Generation

In 3D printing, walls, skins, or shells are the lines of material that follow the object's outer perimeter, providing structural integrity and a smooth surface finish. The outer wall is offset inwards by a distance of half the extrusion width from the cross-section. To generate the inner walls, the outer wall is offset inwards by the extrusion width to create multiple equidistant walls. A robust offsetting method ensures that the walls are properly spaced and do not overlap or leave gaps.

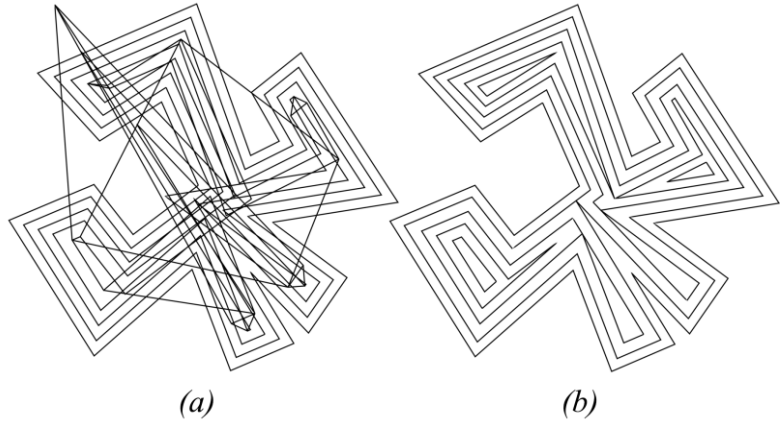

Fig. 9. a) Simple polygonal offset b) Offset by straight skeleton method.

A simple polygon offset algorithm typically shifts the edges of the polygon inwards by a certain distance along the direction perpendicular to the edge and calculates the new intersections [18]. However, the method often fails for complex and dense geometries, resulting in overlapping edges, gaps, and self-intersecting polygons [19, 20]. These artifacts make the resulting geometry unusable for 3D printing applications.

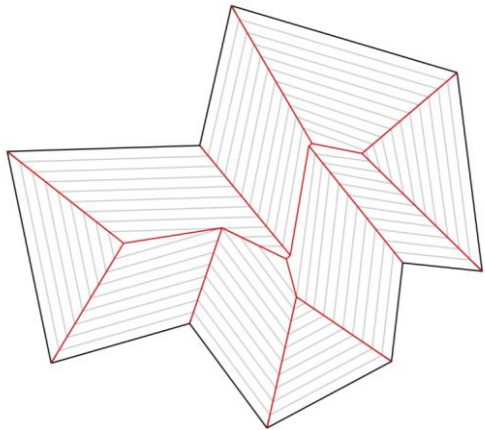

Fig. 10. Offsets produced by the straight skeleton method. In red, are the edge-front interfaces defining the straight skeleton.

The straight skeleton (SS) method produces offsets by a wavefront-propagation process [21, 22]. The SS approach generates polygons by tracing the inward movement of the original polygon's edges moving at unit speed, parallel to the edges. The edges propagate until they meet and form vertices. Points along the edge-front, equidistant from the original polygon, are connected to produce the parallel offsets. This method inherently avoids the creation of overlapping geometry. Using SS, the offset polygons maintain sharp corners and preserve the overall shape, resulting in cleaner offsets free from self-intersections and gaps. This is particularly advantageous for generating a large number of walls that fill the entire area inside the polygon.

## 2.8. Infill Generation

The area inside of the walls is filled following methods that fall into three categories – direction-parallel paths, contour-following paths, and space-filling curves [23]. Due to its simplicity and computational efficiency, direction-parallel paths are commonly used in AM. In FDM type AM, a requirement is to reduce printing time by minimizing the number of times the print head has to retract by following a continuous path as much as possible. So, the print head follows a zig-zag pattern where it does not have to retract between adjacent lines [24]. But for concave polygons i.e. polygons that have at least one concave curve, it is not possible to fill the area with a single continuous zig-zag path. So, polygons are broken down into multiple monotone polygons, and separate zig-zag infill paths are produced for each polygon.

### 2.8.1. Polygon Decomposition

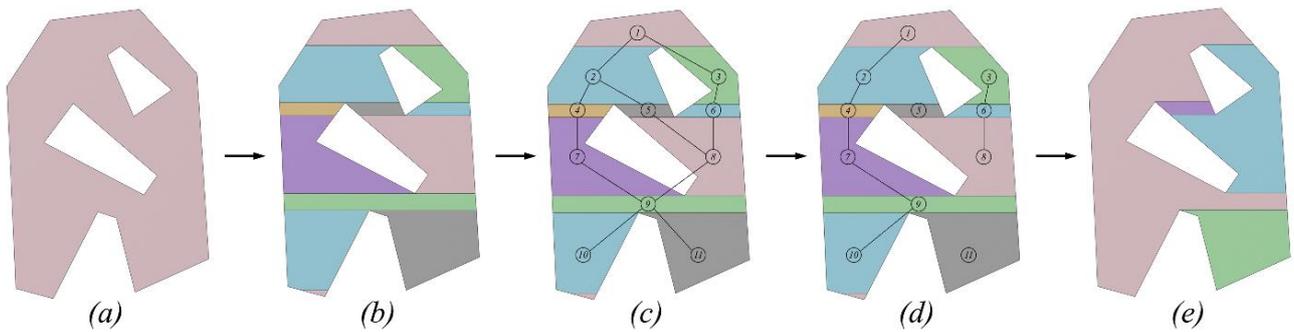

Fig. 11. a) Boundary to generate zigzag infill b) Partitioning polygon by sweep line algorithm c) Creation of adjacency graph d) Creation of different connected components from adjacency graph e) Connecting polygons based on connected components

The boundary for the zigzag path is the last wall generated by the SS method, as mentioned in wall generation. The last wall is not printed but used as the boundary for zig-zag path generation. The first step is to test the polygon for concavity by checking if any internal angle is more than 180 degrees. If the polygon is convex, then decomposition is not needed, and a zigzag path can be generated; otherwise, polygon decomposition is done [25].

In a concave polygon, a merge event occurs when two polygon edges converge inward at a vertex, and a split event occurs when two polygon edges intersect and diverge into separate polygons. The sweep line algorithm identifies the merge and split events [26]. First, the line segments are sorted by their y values. For a zigzag pattern not parallel to the print-bed x-axis, the transformed y-coordinates are sorted. A horizontal line is swept from the top to the bottom of the polygon, maintaining a count of intersections with the polygon edges. The number of intersections above and below the vertex is compared at each point. The points at which the number of intersections changes indicate a merge or split event. The lines passing through the event points partition the polygon into monotone sub-polygons with respect to the sweep line.

The decomposed sub-polygons are merged to form a continuous infill pattern across the entire area of the polygon. First, an adjacency graph is created using the shared partition edges of each monotone polygon, where a number represents each polygon. Connected components through the adjacency graph are determined based on a depth-first graph search (DSF) algorithm such that each monotone sub-polygon is visited only once. Finally, the polygons are connected in the order given by the connected components [27].

**2.8.2. Zig-zag path Generation**

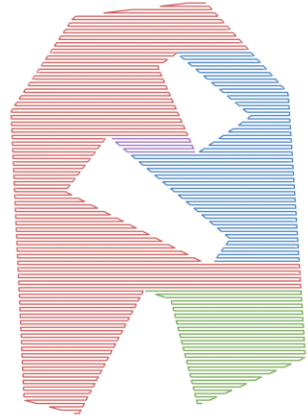

Fig. 12. Generation of different zigzag infill paths in connected polygons.

The connected monotone sub-polygons are now the boundaries to generate the zigzag infill pattern. The incline of the parallel line must be the same as that of the sweep lines. Each line will intersect the polygon at two points, and these points are then connected with the next in alternating directions [28]. The equation for parallel lines is as follows:

$$y = mx + n \times d\sqrt{m^2 + 1} \qquad (8)$$

Here, $d$ represents the perpendicular distance between adjacent lines, $m$ is the slope, and $n$ is the sequence. The infill can be fully dense, with $d$ equal to the extrusion width, or smaller to overlap adjacent infill lines. Alternatively, a user-defined infill percentage may be used, in which case $d$ is greater than the extrusion width, resulting in an infill that consists of alternating lines and voids. Infills are rotated 90 degrees between each layer to increase the strength of the printed object.

**2.9. Non-Planar Extruder Path Generation**

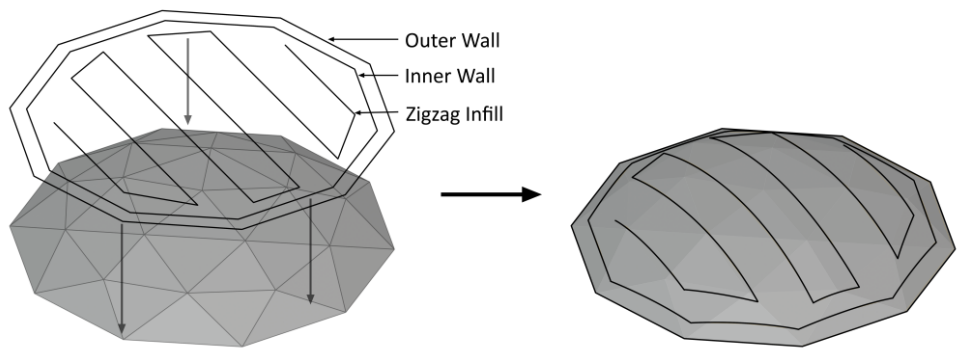

Fig. 13. Generation and projection of non-planar walls and zigzag infill paths onto non-planar surface.

The non-planar extruder paths are produced by first generating the 2D outer wall, inner wall, and zig-zag infill paths, which are then projected onto the non-planar surface to make the conformal 3D paths.

## 2.10. Non-Planar Wall Generation

The non-planar boundary is the 2D projection of triangles of the non-planar surface after discarding the shared edges, i.e., edges that appear more than once. The result is polygonal chains the same as the cross-sections produced from slicing the model and undergoing the same processing. The z-coordinates of the non-planar wall are not defined yet, so they lie flat above the maximum height of the model. The flat polygon is projected onto the offset non-planar surface of the model to match the contour of the model.

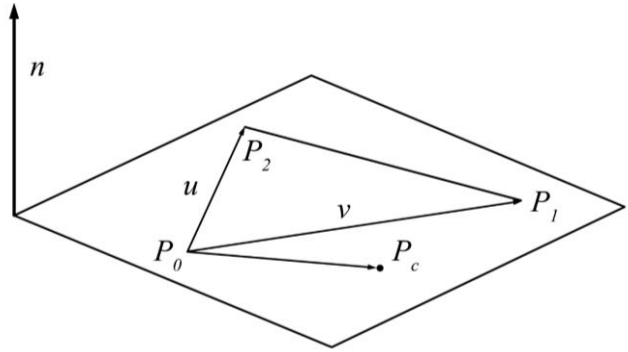

Fig. 14. Calculation of barycentric coordinates.

First, a check is done for the inclusion of the point inside each non-planar triangle by calculating the barycentric coordinates of the point with respect to the triangle. It is adequate to test for inclusivity in terms of x and y-coordinates, reducing the number of calculations. The point and triangle are projected onto a 2D plane and the barycentric coordinates of the point are computed [17]. The barycentric equation is calculated as follows:

$$w = P_c - P_0, u = P_1 - P_0, v = P_2 - P_0 \qquad (9)$$

$$s = \frac{(u.v)(w.v) - (u.v)(w.u)}{(u.v)^2 - (u.u)(v.v)} \qquad (10)$$

$$t = \frac{(u.v)(w.u) - (u.u)(w.v)}{(u.v)^2 - (u.u)(v.v)} \qquad (11)$$

$P_c = V(s,t)$ is inside the triangle $T$ when $s > 0$, $t > 0$, and $s+t < 1$. A threshold value of 0.001 mm was used instead of comparing to zero because it was seen in the tests that some points located exactly at the edges or vertex of triangles were computed outside the triangle due to floating-point precision error.

The z-coordinate of the point is the height the extruder should reach and must be set to the same height the corresponding triangle has in this position [10]. If the condition of inclusivity of the point inside the triangle is satisfied, it is projected along the z direction, and the intersection with the plane of the triangle is calculated using the ray-plane intersection equations used in the slicer. Repeating this for all points of the chain, the z-coordinates of the walls are derived. The extruder orientation for a 5-axis system is the average of facet normal of the triangles the point lies in between [29]. The procedure is discussed in Algorithm 1.

Algorithm 1: Extruder path generation of non-planar walls

**Step 1:** Non-Planar triangles detection

**Input:** Vectors and normals from STL file: *my_mesh.vectors* and *my_mesh.normals*, non-planar threshold angle: *angle_threshold*

**Output:** Non-planar triangles: *nonplanar_vectors*

For i from 0 to length(*my_mesh.vectors*):
    Find the angle between *my_mesh.normals[i]* and the vertical direction *angle*
    If *angle* <= *angle_threshold* and *my_mesh.normals[i][2]* > 0:
        For j from 0 to length(*my_mesh. vectors*):
        Find the barycentric coordinate *u, v, w* of vertices of *my_mesh.vectors[i]* with respect to *my_mesh.vectors[j]*
        If $u + v >= 1.001$ and $u >= -0.001$ and $w >= -0.001$ and *my_mesh.normals[j][2]* > 0:
            Continue
        Else:
            Add *my_mesh.vectors[i]* to *nonplanar_vectors*

**Step 2:** Offset and projection of non-planar boundary

**Input:** Non-planar triangles: *nonplanar_vectors*

**Output:** Non-planar walls extruder paths: *nonplanar_walls*

Delete shared edges in *nonplanar_vectors* and add to *nonplanar_boundary*

Sort *nonplanar_boundary*

If *wall_count* != *number_of_walls*:
    If *wall_count* == 0:
        Offset distance $d = 0.5 * $ *extrusion_width*
    Else:
        Offset distance $d = (0.5 + $ *wall_count*$) * $ *extrusion_width*
    Find offset of *nonplanar_boundary* with offset distance equal to *d* and add to *nonplanar_walls*
    For i from 0 to length(*nonplanar_walls*):
        For j from 0 to length(*nonplanar_vectors*):
            Find the barycentric coordinate *u, v, w* of *nonplanar_walls[i]* with respect to *nonplanar_vectors[j]*
            If $u + v >= 1.001$ and $u >= -0.001$ and $w >= -0.001$:
                Find projection of *nonplanar_walls* on *nonplanar_vectors[j]* and add z-coordinates to *nonplanar_walls*
    *wall_count* += 1

## 2.11. Non-Planar Infill Generation

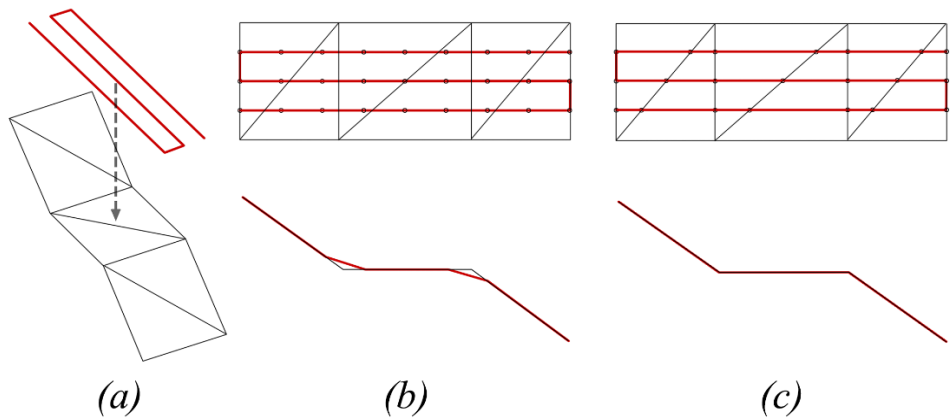

Fig. 15. a) Projection of zigzag path onto surface, in black is the target surface and in red is the projected path b) Subdivision of paths based on constant Euclidean distance c) Subdivision based on intersection of paths with shared edges of 2D projection of non-planar surface.

The paths for non-planar infills are generated using the same zig-zag path generation method as planar infills. Each line consists of a start and end point. However, these two points are inadequate to match the model's contour accurately. Rodriguez et al. [8] and Marie et al. [10] divided each line in the path by equidistant points. Their method allows the setting of a spatial resolution based on Euclidean 2D distance between two adjacent points in the lines. The points are projected onto the model to create a curved path. The smaller the distance between segments, the better the non-planar infills conform to the object's contour. However, this method can result in many points and an increasing number of computations.

Moreover, the points may not align with the edges of the triangles of the non-planar surface, where the contour changes occur, leading to non-conformity. As seen in Figure. 15-b, the projection does not properly match the target surface due to the vertex not aligning with the edges.

A more efficient approach is to add points to the paths only where they intersect with the internal shared edges of triangles of the 2D projection of the non-planar area. This ensures that the interior points will be collinear with the edge where the contour changes can occur. Since the line is subdivided by points only where needed, the projected infill paths more faithfully conform to the target surface's curvature. The path's intersection points with the non-planar surface's shared edges are calculated and sorted in terms of the x-coordinates when the slope is greater than 0. Finally, the points are projected onto the non-planar surfaces to generate curved infill extruder paths. When projecting a 2D zig-zag pattern onto a 3d surface, the large slope on the projecting surface can increase the actual spatial distance between adjacent zig-zag lines. This can increase the under filling in the non-planar surfaces. The problem can be resolved by setting a nonplanar threshold angle up to the acceptable spatial distortion level.

Due to surface offsetting, the boundary of the surface is pulled inwards or outwards, depending on its shape. So, the 2D paths must be recalculated for each surface, projecting only to the corresponding surface. The process is repeated for all the inner non-planar surfaces until the desired number of non-planar layers is reached. The procedure is discussed in Algorithm 2.

Algorithm 2: Extruder path generation of non-planar infills

**Step 1:** Polygon decomposition

**Input:** Non-planar boundary: *nonplanar_boundary*

**Output:** Infill boundary: *polygons*

Offset distance *d* = (0.5 + *number_of_walls*)

Find the boundary of infill by offsetting *nonplanar_boundary* by offset distance *d* to *infill_boundary*

Sort *infill_boundary* in terms of y-coordinate values

For *i* from 0 to length(*infill_boundary*):

    Find the number of intersections of line along vertices of *infill_boundary* with edges of *infill_boundary* and add to *intersection_number*

For i from 1 to length(intersection_number) – 1:

    If *intersection_number[i]* != *intersection_number[i-1]* or *intersection_number[i]* != *intersection_number[i+1]*:

        Add y-coordinate in *infill_boundary[i]* to *partition_edges*

For *i* from 1 to length(*partition_edges*):

    Find vertices above and on *partition_edges[i]* and add to *sub_polygons*

    Delete all vertices above *partition_edges[i]* in *infill_boundary*

Initialize an empty graph *G*

For *i* from 0 to length(*sub_polygons*):

    For *j* from *i*+1 to length(*sub_polygons*):

        Find the number of shared vertices between *sub_polygons[i]* and *sub_polygons[j]* and add to *shared_vertices*

        If *shared_vertices* >= 2:

          Add an edge between *i* and *j* in graph *G*

Initialize empty array *paths*, *connected_components*, and *visited*

For *i* from 0 to length(*sub_polygons*):

  If *visited[i]* == FALSE:

    Add *i* to *connected_ components*

    *visited[i]* == TRUE

    For *j* from 0 to length(*sub_polygons*):

      If *visited[j]* == FALSE and an edge exists between *i* and *j* in graph *G*:

        Add *j* to *connected_ components*

        *visited[j]* = TRUE

    Add *connected_ components* to *paths*

For *i* from 0 to length(*paths*):

  For *j* from 0 to length(*paths[i]*):

    Add *sub_polygons[path[i][j]]* to *polygons*

**Step 2:** Generation and projection of infill paths

**Input:** Infill boundary: *polygons,* non-planar triangles: *my_mesh.vectors*

**Output:** Non-planar infills extruder paths: *infill_paths*

Delete shared edges in *polygons*

For *i* in *polygons*:

  Find minimum and maximum y co-ordinates in *polygons[i] y_min* and *y_max*

  Initialize *y = y_min*

  Initialize *n = 1*

  Find shared edges in *my_mesh.vectors* and add to *shared_edges*

  If *y >= y_min* and *y <= y_max*:

    Find the intersection of line *y = y_min + (extrusion_width * n)* with *polygons[i]*

    If *n % 2 != 0*:

      Add intersection points to *infill_paths*

      Find the intersection between *infills_paths* and *shared_edges* and add to *infill_paths*

    Else:

      Swap intersection points and add to *infill_paths*

      Find the intersection between *infills_paths* and *shared_edges* and add to *infill_paths*

  *y = y_min + (extrusion_width * n)*

  *n += 1*

  For *i* from 0 to length(*infill_paths*):

    For *j* from 0 to length(*nonplanar_vectors*):

      Find the barycentric coordinate *u*, *v*, *w* of *infill_paths[i]* with respect to *nonplanar_vectors[j]*

      If *u + v* >= 1.001 and *u* >= -0.001 and *w* >= -0.001:

        Find projection of *infill_paths* on *nonplanar_vectors[j]* and add z-coordinates to *infill_paths*

Table 1. STL mesh, PrusaSlicer results, and Non-planar algorithm results.

| No. | STL Mesh | PrusaSlicer | Non-planar Algorithm |
|-----|----------|-------------|----------------------|
| 1 | 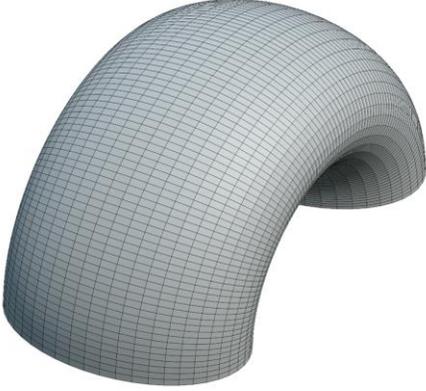 | 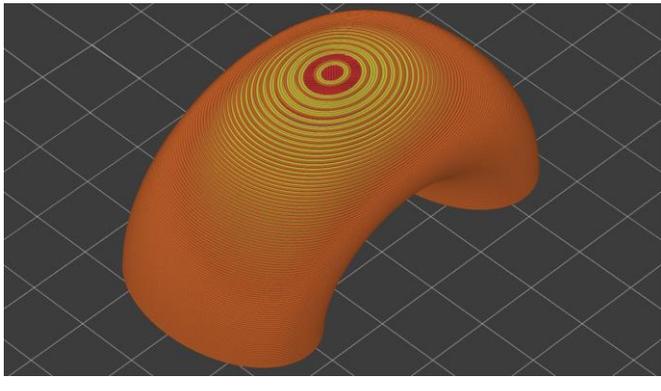 | 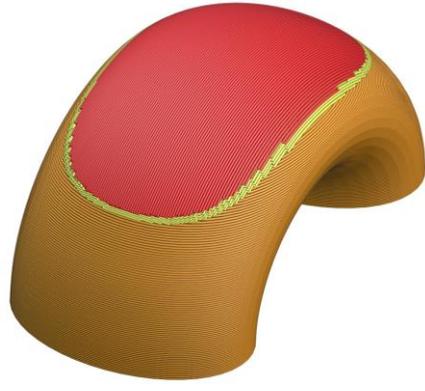 |
| 2 | 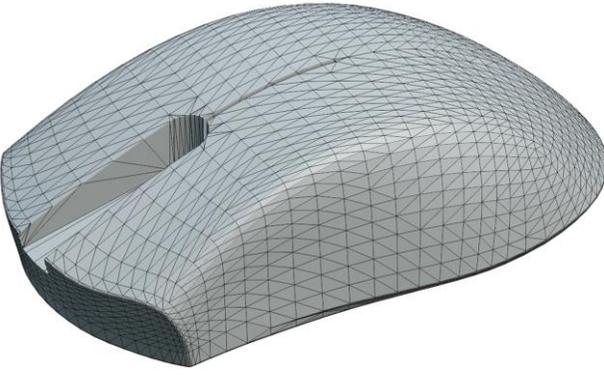 | 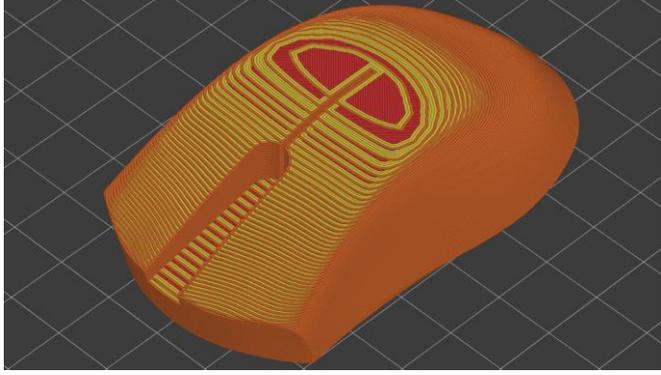 | 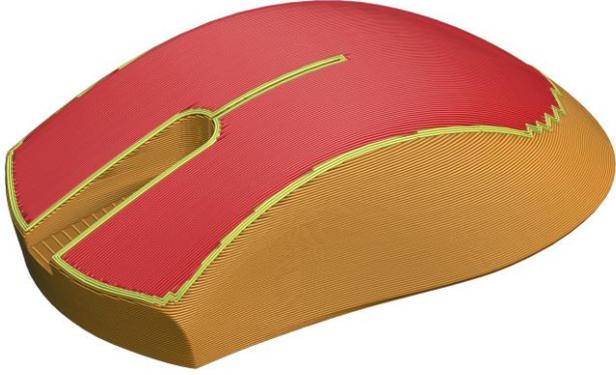 |
| 3 | 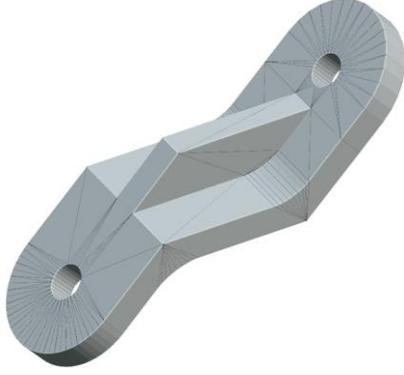 | 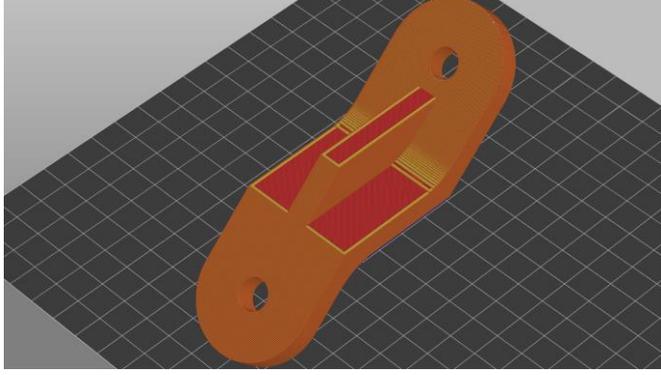 | 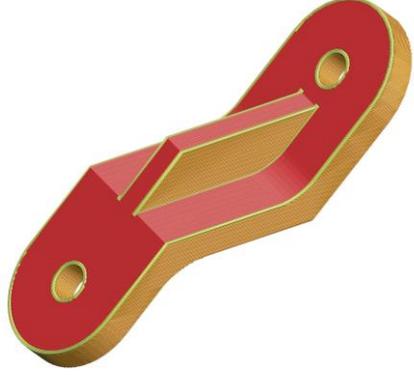 |

Table 1. STL mesh, PrusaSlicer results, and Non-planar algorithm results (continued).

| No. | STL Mesh | PrusaSlicer | Non-planar Algorithm |
|---|---|---|---|
| 4 | 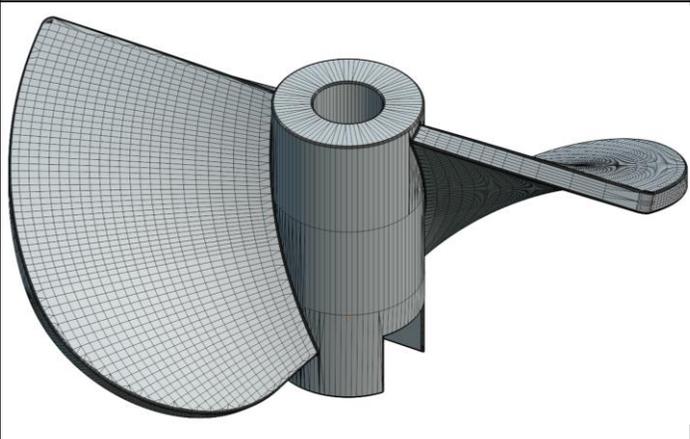 | 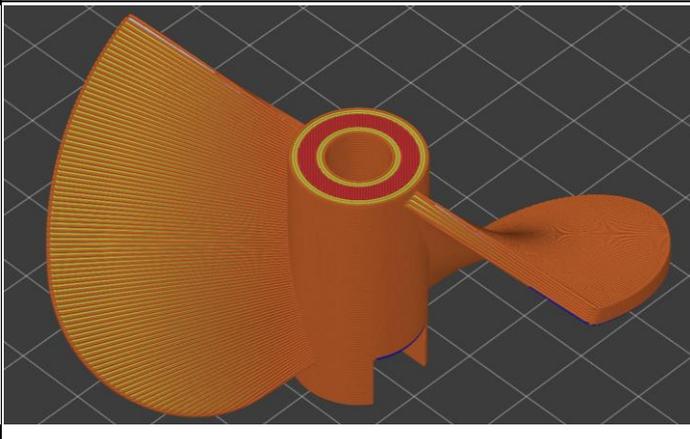 | 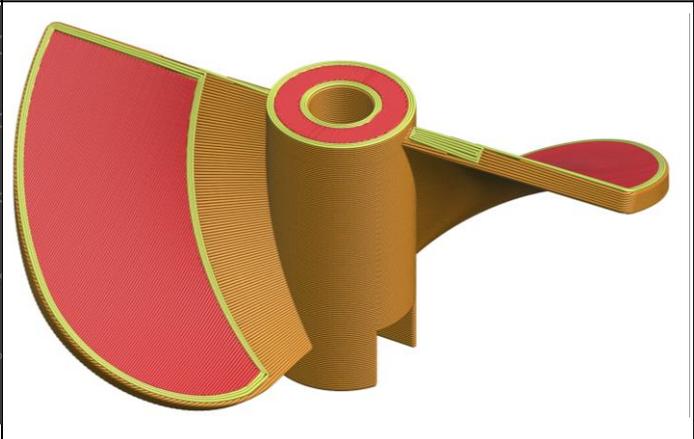 |
| 5 | 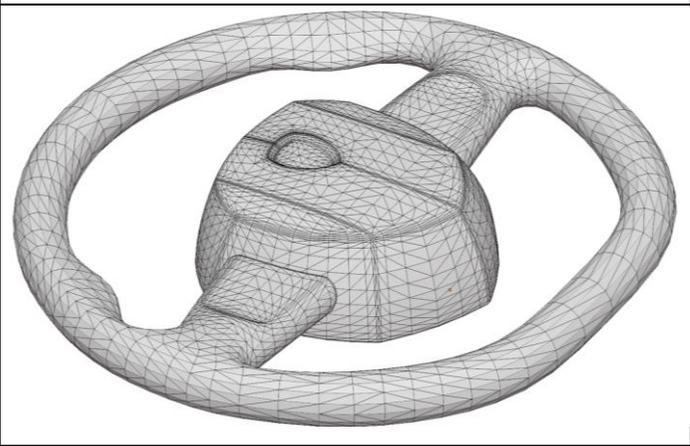 | 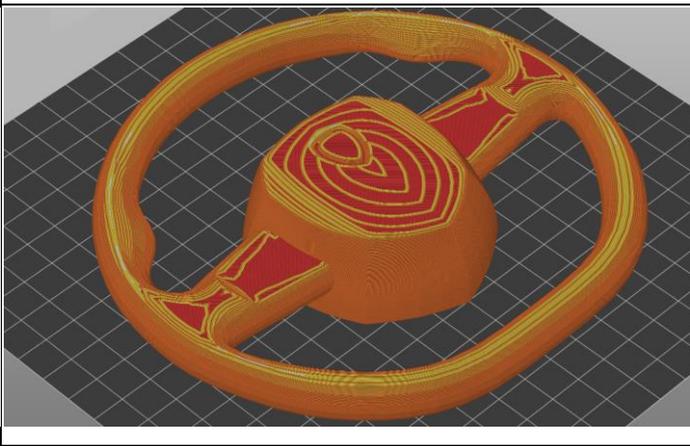 | 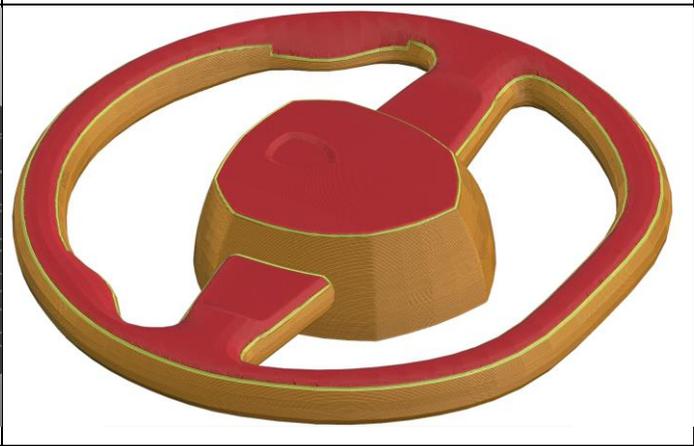 |

# 3. Analysis

## 3.1. Surface accuracy measurement

Table 2. Surface accuracy comparison.

| Original model | Planar output | Nonplanar output |
|---|---|---|
| 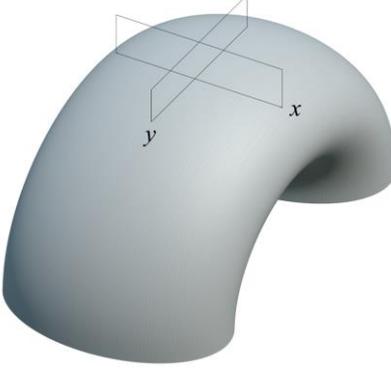 | 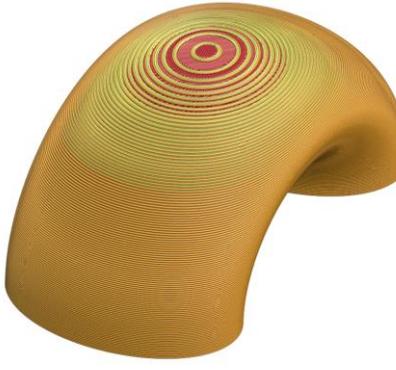 | 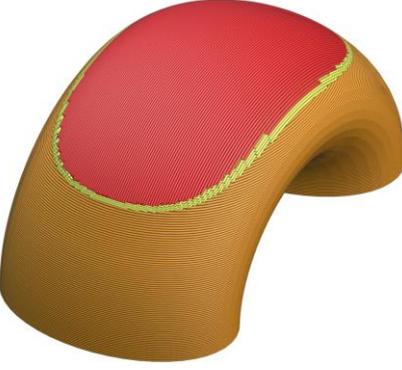 |
| 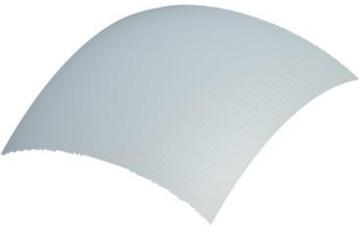 | 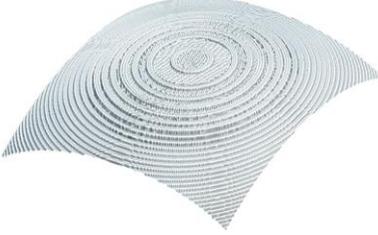 | 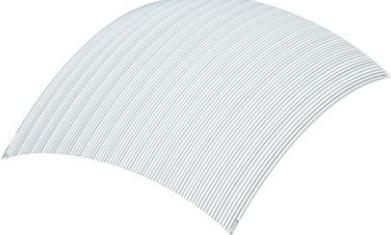 |

To quantify the surface accuracy, the surface of the STL model and the model generated by planar and non-planar extruder paths of the object from Table 1-1 are compared. The threshold angle is set to 90 degrees to generate the planar model, so all surfaces are generated as planar layers. Only the surface at the top of the curved section is compared, where the staircase effect is the most pronounced. A plane containing 124,560 vertices is projected onto the corresponding mesh using the shrink-wrap modifier to get the surface coordinates.

Using the vertex coordinates from the shrink-wrapped plane, the Chamfer Distance (CD) is used to quantify how similar the surfaces are in shape [30]. CD calculates the mean distance from each point in one set to the nearest point in the other and vice versa. P = {$p_1$, $p_2$,…, $p_m$} and Q = {$q_1$, $q_2$,…, $q_n$} are the co-ordinates of the cross-sections, one from the generated models and another from the object's original STL file respectively. The equation to calculate the chamfer distance from set P to Q is as follows-

$$CD(P, Q) = \frac{1}{m}\sum_{i=1}^{m} \min_{j=1} \left\| p_i - q_j \right\| + \frac{1}{n}\sum_{j=1}^{n} \min_{i=1} \left\| q_j - p_i \right\| \quad (12)$$

$\|p_i - q_j\|$ represents the Euclidean distance between point $p_i$ in set P and its nearest neighbor in set Q and vice versa for $\|q_j - p_i\|$. m and n represent the number of points in cross-sections P and Q respectively. The CD of the model generated by planar extruder paths and the original mesh is 0.0864 mm and of the model generated by non-planar extruder paths and original mesh is 0.0112 mm.

Moreover, cross-sections about planes along two axes, one along the hatching direction of the non-planar infill pattern, plane y in Table 2, and the other perpendicular to it, plane x in Table 2, are also shown. The cutting planes are located at the top, where the staircase effect is most pronounced.

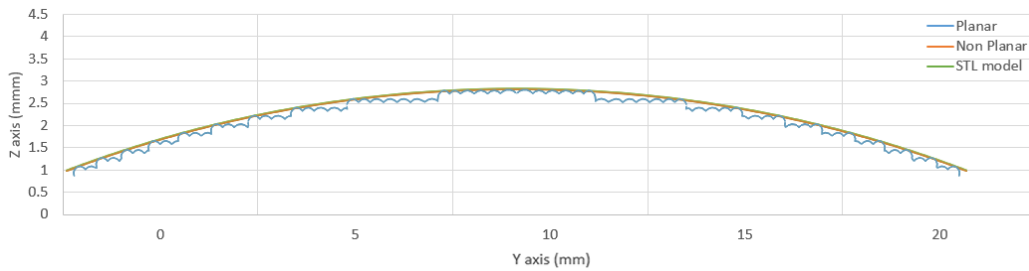

Fig. 16. Cross-section along hatching direction- plane y.

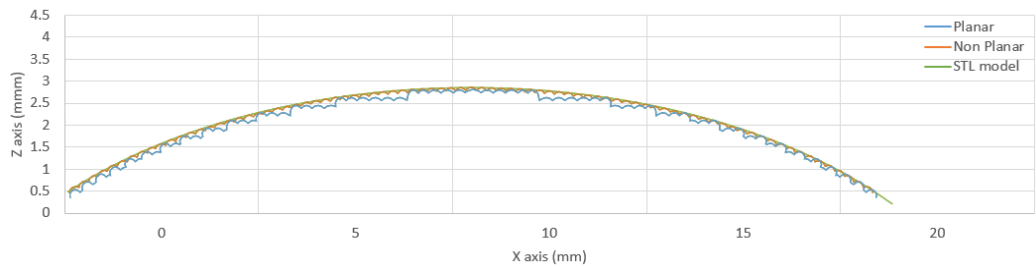

Fig. 17. Cross-sections against hatching direction- plane x.

## 4. Results and Discussion

The extruder paths produced by the algorithm for 5 different cases are presented in Table. 1. All the tests are produced with a layer height of 0.3 mm and an extrusion width of 0.4 mm. The extruder paths are generated by a Python program and produced as a Bezier curve. For graphical simulation, the curve was imported to Blender and given volume using the 3D curve feature. For comparison, the planar slicing results are made using PrusaSlicer.

Table. 1-1 shows a freeform convex surface. Slicing the object by planar methods shows a high amount of stair-stepping, becoming more intense at the top. Meanwhile, slicing the same object with the non-planar algorithm significantly reduces the stair-stepping effect. In this case, the non-planar boundary is not distributed at the same z-height but at varying heights, depending on the shape of the non-planar surface. Despite the irregularities, the algorithm produced a smooth transition between the non-planar and planar layers, which was not demonstrated in previous cases [9, 10]. CD calculations for this model show that the object generated by non-planar extruder paths is 8 times more similar to the actual model than the model generated by planar extruder paths. As seen in the cross-section in Figure X, surface accuracy is dependent on the hatching direction. Cross-sections of the non-planar model taken along the hatching direction show high accuracy, while those taken perpendicular show reduced accuracy due to bumps between adjacent infill lines. This directional sensitivity or anisotropic accuracy, resulting from the alignment from the hatching must be considered when configuring the non-planar slicer.

Table 1-2 is a convex surface similar to the model of Table 1-1, but applied to a practical case: the body of a computer mouse. Polygonal decomposition allows the algorithm to generate non-planar extruder paths of surfaces with splits and holes. The model of Table. 1-3 is a hard surface object with well-defined, sharp edges and flat and angular faces. In this case, three non-planar layers were possible, after which the offset surface starts self-intersecting. The part is particularly challenging because of the rib in the middle, which appears as a zero-thickness hole in the 2D projection of the non-planar surface. The SS method generated the offsets for the non-planar walls and the infill boundary. This object's limitation of the current algorithm is evident because a typical FDM extruder would collide with the planar part of the rib section when printing the non-planar surface in the middle.

Table 1-4 shows a propeller with a high requirement for surface smoothness. Propeller blades have curvature in different directions, and the algorithm produced accurate non-planar extruder paths of these complex surfaces. Table 1-5 shows a steering wheel and demonstrates the algorithm's ability to process multiple surface patches distributed at different z-heights. Significant aesthetic improvements are evident in this case.

In all these cases, in practical applications, a typical FDM extruder would collide with the planar layers when there is more than one non-planar layer. As the numbers of non-planar layers increase, the extruder must travel deeper into the object. However, printing these objects with more than one non-planar layer would be possible if thin extruders are used where the extruder is smaller in diameter than the extrusion width, such as the airbrush nozzles used by Pelzer et al. [31] or the dispensing needles used in direct ink writing printing, and suspension bath printing [32].

## 5. Conclusion

This work presents a non-planar slicing algorithm capable of generating combined planar and non-planar extruder paths for geometrically complex objects. The printed object consists of interior planar layers that also function as supports and top non-planar layers that conform to the object's shape. The resulting extruder paths are intended for use with a 5-axis 3D printer.

The algorithm was tested on various objects with differing complexities and challenges, demonstrating versatility not seen in previously developed methods. Graphical simulations show a significantly reduced staircase effect and improved aesthetic quality and accuracy.

The current study is limited to graphical simulations due to constraints in equipment availability. Future work will focus on implementing mesh-collider collision detection, flow and velocity control, G-code generation, and beneficial nozzle geometry for non-planar printing to allow the 3D printing of actual physical objects.

**Funding Sources**

This research did not receive any specific grant from funding agencies in the public, commercial, or not-for-profit sectors.